\documentclass[preprint,aps,amsmath,superscriptaddress,nofootinbib]{revtex4}

\usepackage{dcolumn}
\usepackage{bm}
\usepackage{epsfig}
\usepackage{graphicx}
\usepackage{amsmath}
\usepackage{amssymb}
\usepackage{mathrsfs}
\usepackage{color}
\usepackage[usenames,dvipsnames]{xcolor}
\usepackage{hyperref}
\usepackage{braket}
\usepackage[caption=false]{subfig}
\usepackage{multirow}
\usepackage{array}
\usepackage{slashed}

\usepackage{amsfonts}

\def\be{\begin{equation}}
\def\ee{\end{equation}}
\def\bea{\begin{eqnarray}}
\def\eea{\end{eqnarray}}

\newcommand{\pv}{{\bf p}}
\newcommand{\Pv}{{\bf P}}
\newcommand{\qv}{{\bf q}}

\newcommand{\kv}{{\bf k}}
\newcommand{\bv}{{\bf b}}

\def\slashchar#1{\setbox0=\hbox{$#1$}           
   \dimen0=\wd0                                 
   \setbox1=\hbox{/} \dimen1=\wd1               
   \ifdim\dimen0>\dimen1                        
      \rlap{\hbox to \dimen0{\hfil/\hfil}}      
      #1                                        
   \else                                        
      \rlap{\hbox to \dimen1{\hfil$#1$\hfil}}   
      /                                         
   \fi}

\begin{document}

\title{Polarized TMD fragmentation functions for $J/\psi$ production}


\author{Marston Copeland}
\email{paul.copeland@duke.edu}
\affiliation{Department of Physics, Duke University, Durham, North Carolina\ 27708, USA\\}

\author{Sean Fleming}
\email{spf@email.arizona.edu}
\affiliation{Department of Physics and Astronomy, University of Arizona, Tucson, Arizona\ 85721, USA\\}

\author{Rohit Gupta}
\email{rohitkgupta@arizona.edu}
\affiliation{Department of Physics and Astronomy, University of Arizona, Tucson, Arizona\ 85721, USA\\}

\author{Reed Hodges}
\email{reed.hodges@duke.edu}
\affiliation{Department of Physics, Duke University, Durham, North Carolina\ 27708, USA\\}

\author{Thomas Mehen}
\email{mehen@duke.edu}
\affiliation{Department of Physics, Duke University, Durham, North Carolina\ 27708, USA\\}

\begin{abstract} 
We calculate the matching, at leading order, of the transverse momentum-dependent fragmentation functions (TMDFFs) for light quarks and gluons fragmenting to a $J/\psi$ onto polarized nonrelativistic QCD (NRQCD) TMDFFs. The NRQCD TMDFFs have an operator-product-expansion in terms of nonperturbative NRQCD production matrix elements. Using the results we obtain, we make predictions for the light quark fragmentation contribution to the production of polarized $J/\psi$ in semi-inclusive deep inelastic scattering (SIDIS) both for unpolarized and longitudinally polarized beams. These results are an important contribution to polarized $J/\psi$ production in SIDIS, and thus are needed for comparison with experiments at the future Electron-Ion Collider.
\end{abstract}

\maketitle

\section{Introduction}


The successful application of the parton model to QCD over the past few decades has offered a rich understanding of hadron structure, despite the theory's nonperturbative nature. In this picture, constituent quarks and gluons can be identified with partons, allowing for factorization theorems of high-energy cross sections like those for $e^+ e^-$ annihilation into hadrons \cite{Collins:1981va}, Drell-Yan \cite{Drell:1970wh}, and semi-inclusive deep inelastic scattering (SIDIS) \cite{Feynman:1969ej,Bloom:1969kc,Ji:2004wu}. These theorems separate strong processes into universal, process-independent, nonperturbative structures and hard, process-dependent, perturbative cross sections for partonic scattering \cite{Collins:1989gx}. The nonperturbative parts are encoded by the parton distributions, such as parton distribution functions (PDFs) or fragmentation functions (FFs), which can be extracted independently \cite{Cammarota:2020qcw} and are universal. 

When the kinematics are sensitive to both the longitudinal momentum fraction of partons and the small transverse momentum relative to their bound state, collinear parton distributions can be generalized to transverse momentum-dependent (TMD) distributions \cite{Angeles-Martinez:2015sea,Boussarie:2023izj}. TMD PDFs and FFs have attracted great attention over the past few years because of their promise to give insight into the 3D structure of hadrons. They can provide unique information on the dynamics of polarized parton constituents with respect to polarized hadron observables. This makes them particularly appealing for experiments with polarized beams or targets, like those proposed for the upcoming Electron-Ion Collider (EIC) \cite{Accardi:2012qut, AbdulKhalek:2021gbh}.

Like their collinear counterparts, TMD PDFs and FFs are still nonperturbative and have to be extracted from experiment or calculated on the lattice. Precise measurements of TMD functions greatly benefit from experimental observables that are clean. Since fragmentation requires an identified hadron in the final state, one strategy for obtaining a clear signal is to single out hadrons with clean decay channels. The $J/\psi$ meson, which has historically played an important role in our understanding of the strong interactions, has a particularly clean signal in its decay to two leptons (roughly 14\% of the time).

The $J/\psi$ is a doubly heavy bound state of a charm and anticharm quark, and falls into the larger category of quarkonium states (bound states of a heavy quark and heavy antiquark). Theoretically quarkonium is especially attractive because the large masses of the heavy quarks allow the application of nonrelativistic QCD (NRQCD) \cite{Bodwin:1994jh,Luke:1999kz,Brambilla:1999xf}. 
%
The success of NRQCD when applied to collinear FFs of quarkonium suggests that the same can be done for their TMD counterparts, however this subject remains almost untouched. In addition to giving insight into the structure of probed hadrons, studying TMDFFs for quarkonium like the $J / \psi$ also offers more avenues to test the NRQCD factorization conjecture, which has recently come under scrutiny from global fits of long-distance NRQCD matrix elements to  world data \cite{Butenschoen:2011yh,Butenschoen:2012qr, Bodwin:2014gia, Chao:2012iv}.

Prior work \cite{Lee:2021oqr,Catani:2014qha,Ma:2014svb,Kang:2014tta,Sun:2012vc,Catani:2010pd,Mukherjee:2016cjw,Mukherjee:2015smo,Boer:2012bt,Echevarria:2019ynx,Fleming:2019pzj,DAlesio:2021yws,Boer:2020bbd,Bor:2022fga,Kishore:2021vsm,Scarpa:2019fol,DAlesio:2019qpk,Bacchetta:2018ivt,Mukherjee:2016qxa,Rajesh:2018qks,Godbole:2013bca,Godbole:2012bx,denDunnen:2014kjo,Kang:2014pya,Zhu:2013yxa} exists in the literature on transverse momentum dependence in quarkonium direct production processes. Azimuthal asymmetries, in particular, have been studied in great detail \cite{Bor:2022fga,DAlesio:2021yws,Kishore:2021vsm,Boer:2020bbd,Scarpa:2019fol,DAlesio:2019qpk,Bacchetta:2018ivt,Mukherjee:2016qxa}, as well as single-spin asymmetries \cite{Rajesh:2018qks,Godbole:2013bca,Godbole:2012bx}. Some authors \cite{DAlesio:2021yws,Kang:2014pya} have considered $J/\psi$ polarization in their analysis. References \cite{Echevarria:2019ynx,Fleming:2019pzj,Boer:2020bbd,DAlesio:2021yws} investigate the effects of including TMD shape functions when studying quarkonium production. A significant motivation for studying quarkonium production has been its potential as an avenue to probe gluon TMDs 
\cite{Godbole:2012bx,Boer:2012bt,Godbole:2013bca,denDunnen:2014kjo,Mukherjee:2015smo,Mukherjee:2016cjw,Mukherjee:2016qxa,Rajesh:2018qks,Scarpa:2019fol,DAlesio:2019qpk,Kishore:2021vsm,Zhu:2013yxa}. However, nearly all of these studies, with the exception of \cite{Echevarria:2020qjk}, neglect TMD fragmentation which gives contributions that are comparable, if not greater, in magnitude than $J/\psi$ production from photon-gluon fusion.  Beyond the realm of quarkonium, other authors have studied TMDFFs for heavy quarks fragmenting to heavy hadrons \cite{vonKuk:2023jfd,Dai:2023rvd}.

In this paper we calculate the polarized TMDFFs for quarks and gluons fragmenting to a $J/\psi$ final state for the first time. In Sec.~\ref{sec: defs} we define the kinematics of the relevant processes, review the TMD fragmentation functions, and review the NRQCD factorization conjecture. In Sec.~\ref{sec: q fragmentation} we perturbatively calculate the short distance coefficients of TMDFFs for quarks fragmenting into a $J/\psi$ hadron at leading order in $\alpha_s$. Then, we consider the various combinations of quark and $J/\psi$ polarizations to project out all possible polarized TMDFFs. In Sec.~\ref{sec: g fragmentation} we repeat these steps for a gluon fragmenting to a $J/\psi$. Finally, in Sec.~\ref{sec: phenom} we study the polarized fragmentation functions numerically and use these results to make predictions for the $^3S_1^{[8]}$ $J/\psi$ fragmentation contribution to polarized SIDIS cross sections. 

\section{Definitions}
\label{sec: defs}
\subsection{Notation}


We first begin with a quick overview of the notation we use. Our light-cone coordinate vectors $n$ and $\bar{n}$ are defined as:
\begin{equation}
    \begin{aligned}
        n^\mu = & \; \frac{1}{\sqrt{2}}(1,0,0,-1) \, , \\
        \bar{n}^\mu = & \; \frac{1}{\sqrt{2}}(1,0,0,1) \, .
    \end{aligned}
\end{equation}
Any vector $v$ can be decomposed as $v^\mu = (n\cdot v)\bar{n}^\mu + (\bar{n}\cdot v)n^\mu + v_T^\mu$.  Notationally, $ n\cdot v \equiv v^+$ and $ \bar{n}\cdot v \equiv v^-$. The transverse components of a vector can be projected out with $g_T^{\mu \nu} \equiv g^{\mu \nu}-n^{\mu} \bar{n}^{\nu}-n^{\nu} \bar{n}^{\mu}$ such that $g_T^{\mu \nu}v_\nu=v_T^\mu$. 

For fragmentation we work in a frame where the $J/\psi$ momentum is
\begin{equation}
    \begin{aligned}
       P^\mu = & \; P^+ \bar{n}^\mu + \frac{M^2}{2P^+}n^\mu \; .
    \end{aligned}
\end{equation}
This has no transverse momentum, i.e. $g_T^{\mu \nu}P_\nu=0$, so the transverse momentum dependence of the fragmentation functions will be described only by the transverse momentum of the fragmenting partons. The TMD fragmentation functions will, in general, be dependent on $z$, the fraction of longitudinal momentum the $J/\psi$ inherits from the fragmenting parton. 

%

\subsection{TMDs}
The bare transverse momentum-dependent fragmentation function for a quark  of flavor $i$ to fragment into $J/\psi+X$, where $X$ represents other possible particles in the final state, is defined as \cite{Boussarie:2023izj}
\begin{equation}
\begin{aligned}
\label{eq: q TMDFF}
\tilde{\Delta}^{(0)}_{q\to J/\psi} (z, \bv_T, P^+/z) =& \; \frac{1}{2 z N_c}{\rm Tr}\int \frac{d b^-}{2 \pi}e^{ib^-P^+/z}\sum_X \Gamma_{\alpha \alpha'} \\
& \times \bra{0} W_\lrcorner \psi_i^{(0), \alpha}(b) \ket{J/\psi (P) , X}  \bra{J/\psi (P), X} \bar{\psi}_i^{(0),\alpha'} (0) W_\urcorner \ket{0}.
\end{aligned}
\end{equation}
Here and throughout this paper, boldface indicates a Cartesian three-vector. The position vector $b$ has no plus component, $b = (b^-, 0, \bv_T)$. The zero superscript indices indicate bare quantities. The trace is over spin and color indices and $\Gamma \in \{\gamma^+/2, \gamma^+\gamma_5/2, i\sigma^{\beta +}\gamma_5/2\}$ covers the Dirac structures that project out the polarizations of the quark at leading twist.  The half staple shaped Wilson lines are defined as 
\begin{equation}
\begin{aligned}
    W_\lrcorner = & \; W_{\hat{b}_T}(+\infty n;b_T,+\infty) W_{n}(b;0,+\infty) \\
    W_\urcorner = & \; W_{n}(0;+\infty,0) W_{\hat{b}_T}(+\infty n;+\infty,0) \,.
\end{aligned}
\end{equation}
The Wilson line along a generic path $\gamma$ is defined by the path-ordered exponential
\begin{align}
	W_\gamma
	= {\cal P} \exp\biggl[ -i g_0 \int_{\gamma}  dx^\mu A^{c, 0}_\mu(x)\, t^c \biggr]
	\,.
\end{align}
Thus the usual lightlike Wilson line is
\begin{equation}
    W_n (x^\mu; a, b) =  {\cal P} \, {\rm exp}\left[-i g_0 \int_a^b ds~ n\cdot A^{c, 0} (x + s n) t^c\right].
\end{equation}

Similarly, the bare TMDFF for a gluon fragmenting to a $J/\psi$ is defined as \cite{Boussarie:2023izj}
\begin{equation}
\begin{aligned}
\tilde{\Delta}^{(0), \alpha \alpha'}_{ g \rightarrow J/\psi}(z, \bv_T, P^+z ) = \frac{1}{2 z^2 P^+} \int\frac{d b^-}{2 \pi} & e^{i b^- P^+/z} \sum_X  \bra{0} G^{(0), + \alpha}(b) {\cal W}_\lrcorner \ket{J/\psi(P), X} \\ 
\times & \bra{J/\psi(P), X} G^{(0), + \alpha'}(0)  {\cal W}_\urcorner \ket {0}.
\end{aligned}
\label{eq: g TMDFF}
\end{equation}\\
where $G^{\alpha \beta}(b)$ is the gluon field strength tensor.  The script $\mathcal{W}$ indicates the color matrices are in the adjoint representation.

%

\subsection{NRQCD factorization formalism}

\begin{figure*}
    \includegraphics[trim=5.2cm 21.0cm 9.1cm 4cm,clip,scale=1]{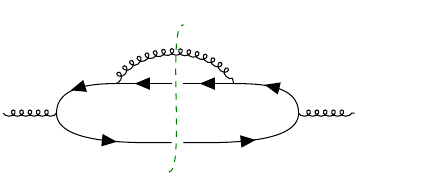}
    \caption{A gluon exchange in a $Q\bar{Q}$ production diagram.}
    \label{fig: gluon across cut}
\end{figure*}

We match the full TMDFFs onto NRQCD using the NRQCD factorization conjecture,
\begin{equation}
    \begin{aligned}
        \Delta_{i\rightarrow J/\psi}(z,\kv_\perp) \rightarrow \sum_{L,s,c,m} d_{i\rightarrow c\bar{c}}^{(m)}(z,\kv_\perp) \braket{\mathcal{O}^{J/\psi}(^{2s+1}L_J^{[c]})} \, ,
    \end{aligned}
\label{eq: NRQCD fact}
\end{equation}
where $d_{i\rightarrow c\bar{c}}^{(m)}(z,\kv_\perp)$ is the hard matching coefficient describing the production of a $c\bar{c}$ from a light parton $i$ and $\braket{\mathcal{O}^{J/\psi}(^{2s+1}L_J^{[c]})}$ is an NRQCD long-distance matrix element (LDME) that describes the hadronization of a $c\bar{c}$ in an angular momentum and color state $^{2s+1}L_J^{[c]}$ into a $J/\psi$. The index $m$ denotes the perturbative order of the matching coefficient.

Equation (\ref{eq: NRQCD fact}) has been applied before to TMD quarkonium production \cite{Echevarria:2020qjk}, however, this is an approximation. For $J/\psi$ transverse momentum below the hadronic scale (approximately 1 GeV) the NRQCD LDME should be replaced with a more sophisticated operator that depends on transverse momentum. To demonstrate why, consider the emission of soft gluons by the $c\bar{c}$, with a single emission shown in Fig.~\ref{fig: gluon across cut}. Here, the soft gluon has momentum of order $m_c v$, where $v$ is the small relative velocity of the $c\bar{c}$ pair. The collinear momentum of the $c\bar{c}$ is $P^+ \gg m_c$, so it is hardly changed by the emission of the gluon. This implies the $J/\psi$'s collinear momentum is the same as the $c\bar{c}$'s and does not change during hadronization. Hence, the NRQCD operators are independent of the longitudinal momentum fraction, $z$. On the other hand, a soft gluon emitted from the charm quark can change the quark's momentum by ${\cal O}(m_c v)$, and a second soft gluon emitted from the anticharm quark can give a configuration where the relative momentum of the $c$ and $\bar{c}$ is zero, but the total momentum of the $c\bar{c}$ pair differs from the $J/\psi$ momentum by ${\cal O}(m_c v)$. As a result, in the small transverse momentum regime ($p_\perp \sim m_c v \sim 1$ GeV) the $c\bar{c}$ produced in the hard interaction can have  substantially different transverse momentum compared to the $J/\psi$. In this picture the LDME would be replaced by a nonlocal NRQCD operator with a small transverse separation, 
\begin{equation}
    \begin{aligned}
        \braket{\mathcal{O}^{J/\psi}(^{2s+1}L_J^{[c]})}  & \to \braket{\mathcal{O}^{J/\psi}(^{2s+1}L_J^{[c]})(\boldsymbol{k}_\perp)} = \int \frac{d^2\bv_\perp}{(2\pi)^2} e^{-i\bv_\perp \cdot \kv_\perp} \\
        & \times \bra{0}\chi^\dagger(\bv_\perp) S_v(\bv_\perp)\mathcal{K} S^\dagger_v(\bv_\perp) \psi(\bv_\perp) \mathcal{P}^{J/\psi} \psi^\dagger(0) S_v(0) \mathcal{K} S^\dagger_v(0) \chi(0) \ket{0},
    \end{aligned}
\end{equation}
where the soft Wilson lines, $S_v$, in the operators maintain gauge invariance and ensure the correct infrared behavior of NRQCD~\cite{Nayak:2005rw,Nayak:2005rt,Nayak:2006fm}. It is the soft gluons in these Wilson lines that are responsible for shifting the total transverse momentum of the $c\bar{c}$ pair.
This NRQCD matrix element is no longer identically an NRQCD LDME; it has a transverse momentum dependence. Similar nonlocal NRQCD operators have appeared in the literature in the form of TMD shape functions during the study of quarkonia P-wave decay to light quarks~\cite{Fleming:2019pzj}. 

The transverse momentum dependence of the NRQCD operator results in a slightly more complicated matching condition than the one presented in Eq.~(\ref{eq: NRQCD fact}),
\begin{equation}
    \begin{aligned}
        \Delta_{i\rightarrow J/\psi}(z,\kv_\perp) \rightarrow \sum_{L,s,c,m} \int \frac{d^2 \qv_\perp}{(2\pi)^2} \, d_{i\rightarrow c\bar{c}}^{(m)}(z,\kv_\perp-\qv_\perp) \braket{\mathcal{O}^{J/\psi}(^{2s+1}L_J^{[c]})(\qv_\perp)} \, .
    \end{aligned}
\label{eq:matchingNRQCD}
\end{equation}

Equation (\ref{eq: NRQCD fact}) is recovered from Eq.~(\ref{eq:matchingNRQCD}) when $\bv^{-1}_\perp \gg m_c v$. In this limit the NRQCD matrix element can be expanded in powers of $\bv_\perp m_c v \ll 1$. At leading order in this expansion,
\begin{equation}
    \begin{aligned}
        \braket{\mathcal{O}^{J/\psi}(^{2s+1}L_J^{[c]})(\boldsymbol{k}_\perp)}  \to &\int \frac{d^2\bv_\perp}{(2\pi)^2} e^{-i\bv_\perp \cdot \kv_\perp} \\
        &\times \bra{0}\chi^\dagger(0) S_v(0)\mathcal{K} S^\dagger_v(0) \psi(0) \mathcal{P}^{J/\psi} \psi^\dagger(0) S_v(0) \mathcal{K} S^\dagger_v(0) \chi(0) \ket{0}\\
        &+ {\cal O}(\bv_\perp m_c v)\\
        =~&\delta^{(2)}(\kv_\perp)\braket{\mathcal{O}^{J/\psi}(^{2s+1}L_J^{[c]})} + {\cal O}(\bv_\perp m_c v)\,.
    \end{aligned}
\label{eq:TMDFFOPE}
\end{equation}
For the purposes of this paper we operate with Eq.~(\ref{eq: NRQCD fact}) as a first-order approximation and save a full treatment of  factorization involving TMD NRQCD matrix elements for another publication \cite{Copeland:2024}.

\section{Quark fragmentation}

\label{sec: q fragmentation}

\begin{figure*}[t]
\centering
\begin{minipage}{0.33\textwidth}
\centering
\subfloat[]{\includegraphics[trim=5.2cm 19.5cm 9.1cm 4.4cm,clip,scale=0.7]{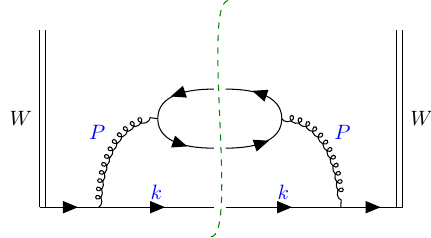}\label{quarkfig1}}
\end{minipage}%
\begin{minipage}{0.33\textwidth}
\centering
\subfloat[]{\includegraphics[trim=5.2cm 19.5cm 9.1cm 4.4cm,clip,scale=0.7]{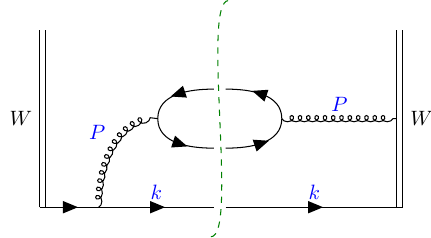}\label{quarkfig2}}
\end{minipage}%
\begin{minipage}{0.33\textwidth}
\centering
\subfloat[]{\includegraphics[trim=5.2cm 19.5cm 9.1cm 4.4cm,clip,scale=0.7]{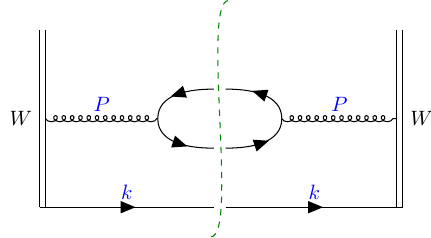}\label{quarkfig3}}
\end{minipage}
\caption{Tree-level diagrams contributing to the quark TMDFF. The double lines represent the Wilson lines, and the single lines are quark propagators. The dashed line represents a cut of the diagram. The mirror diagram of (b) which needs to be included, is not shown.  Momenta are labeled in blue.}
\label{fig: QuarkFeynDiag}
\end{figure*}

As mentioned previously, the short distance coefficients $d^{(m)}_{i\to c\bar{c}}(z,\boldsymbol{p})$ are calculated in QCD perturbation theory. The leading order in $\alpha_s$ contribution to light quark fragmentation is given by the three diagrams [plus mirror image of (b)] shown in Fig. \ref{fig: QuarkFeynDiag}. The amplitudes corresponding to the diagrams are
\begin{equation}
\begin{aligned}
d_A =& \; \frac{g^4}{4z M^4 N_c}\int\frac{d^Dk}{(2\pi)^D} \int db^- e^{ib^-P^+/z}e^{-ib(k+P)}\\
&\times{\rm Tr}\left[\slashed{k}\gamma^\mu \frac{\slashed{k} + \slashed{P}}{(k+P)^2+i\epsilon} \left(\Gamma  \chi_{\mu \nu}\right) \frac{\slashed{k} + \slashed{P}}{(k+P)^2+i\epsilon} \gamma^\nu\right] \delta(k^2)\\
\end{aligned}
\label{DA}
\end{equation}
\begin{equation}
\begin{aligned}
d_{B + {\rm mirror}} = & \; \frac{g^4}{4z M^4 N_c}\int\frac{d^Dk}{(2\pi)^D} \int db^- e^{ib^-P^+/z}e^{-ib(k+P)}\\
&\times{\rm Tr}\left[\slashed{k}\left(\frac{n^\mu}{P^+-i\epsilon}\Gamma \chi_{\mu\nu} \frac{\slashed{k} + \slashed{P}}{(k+P)^2+i\epsilon}\gamma^\nu - \gamma^\mu \frac{\slashed{k} + \slashed{P}}{(k+P)^2+i\epsilon} \Gamma \chi_{\mu\nu}\frac{n^\nu}{P^++i\epsilon}\right)\right]\delta(k^2)\\
\end{aligned}
\label{DB}
\end{equation}
\begin{equation}
\begin{aligned}
d_C =& \; \frac{g^4}{4z M^4 N_c}\int\frac{d^Dk}{(2\pi)^D} \int db^- e^{ib^-P^+/z}e^{-ib(k+P)}\\
&\times{\rm Tr}\left[\slashed{k}\frac{n^\mu}{P^+-i\epsilon}\left(\Gamma  \chi_{\mu \nu}\right) \frac{n^\nu}{P^++i\epsilon}\right] \delta(k^2)\\
\end{aligned}
\label{DC}
\end{equation}
where we have taken the light quark parton mass to be zero, and the fragmenting gluon to have off shellness sufficient to produce a $J/\psi$ of mass $M$ ($P^2 = M^2$). The factor of $\Gamma$ is the Dirac structure required to project out the desired polarization of the parton; the case of an unpolarized parton ($\Gamma=\gamma^+/2$) was done in Ref. \cite{Echevarria:2020qjk}. The $\chi_{\mu \nu}$ corresponds to the spinor structure for the $c \bar{c}$ pair,
\begin{equation}
\begin{aligned}
    \chi_{\mu\nu} = & \; \bar{u}(p) \gamma_\mu T^a v(p') \bar{v}(p')\gamma_\nu T^a u(p). \\
\label{eq: chi}
\end{aligned}
\end{equation}

The matching procedure onto NRQCD involves performing a nonrelativistic expansion of the amplitudes given above. The momentum of the charm quark is  $p=\frac{P}{2} + \Lambda \qv$ and that of the  anticharm quark is $p'=\frac{P}{2} - \Lambda \qv$, where $P$ is the total momentum of the $c\bar{c}$ pair and $\qv$ the relative momentum of the $c$ and $\bar{c}$ in the $c\bar{c}$ rest frame. ${\Lambda^\mu}_i$ is a boost matrix which boosts from the $c\bar{c}$ rest frame to the frame in which the pair has momentum $P$ \cite{Braaten:1996jt}:
\begin{equation}
    \begin{aligned}
        {\Lambda^0}_i &= \frac{1}{2E_\qv}\Pv_i \, , \\
        {\Lambda^i}_j &= {\delta^i}_j - \frac{\Pv^i\Pv_j}{\Pv^2} + \frac{P^0}{2E_\qv} \frac{\Pv^i \Pv_j}{\Pv^2}
    \end{aligned}
\end{equation}

We expand all the kinematic factors in the amplitudes and also need to expand the spinors in Eq.~(\ref{eq: chi}).
Using the identities in \cite{Braaten:1996jt}, Eq.~(\ref{eq: chi}) can be easily expanded to
\begin{equation}
\begin{aligned}
    \chi_{\mu\nu} \approx& \; M^2 {\Lambda^{\mu}}_i {\Lambda^{\nu}}_j  [\xi^\dagger \sigma^i T^a\eta][ \eta^\dagger \sigma^j T^a \xi]. \\
\label{eq: NR spinors}
\end{aligned}
\end{equation}
From here, the nonrelativistic spinor factor $\xi^\dagger \sigma^i T^a\eta \eta^\dagger \sigma^j T^a \xi$ is matched onto the LDMEs with different helicities. 

\subsection{$J/\psi$ polarizations}
\label{sec: Jpsi pol}

 The spinor factor in~(\ref{eq: NR spinors}) matches onto the matrix elements of a NRQCD operator. This operator is written in terms of the NRQCD heavy quark and antiquark fields $\psi$ and $\chi$, as well as a projection $\mathcal{P}_{J/\psi(\lambda)}$ which projects out the $J/\psi$ state of helicity $\lambda$. 
\begin{equation}
    \begin{aligned}
        M^2 \eta^{\prime\dagger} \sigma_i T^a \xi^\prime \xi^\dagger \sigma_j T^a \eta &\leftrightarrow \braket{\chi^\dagger \sigma_i T^a \psi \, \mathcal{P}_{J/\psi (\lambda)} \, \psi^\dagger \sigma_j T^a \chi}  \, , \\
    \end{aligned}
\end{equation}
For the spin-triplet matrix elements for the $J/\psi$, which is an $S$-wave state, spin symmetry implies that the matrix elements must be proportional to $\epsilon^*_{\lambda i} \epsilon_{j \lambda}$ \cite{Braaten:1996jt}, where $\epsilon_{\lambda i}$ is the polarization vector of the $J/\psi$ and $\lambda$ denotes the helicity which can be $\lambda=+1$, $0$, and $-1$. Traditionally, the polarization vectors for the $J/\psi$ are defined by projecting the spin states along the direction of the $J/\psi$'s motion.
%
\begin{equation}
    \begin{aligned}
        \braket{\chi^\dagger \sigma_i T^a \psi \, \mathcal{P}_{J/\psi (\lambda)} \, \psi^\dagger \sigma_j  T^a \chi} &= \frac{2M}{3} \epsilon^*_{\lambda i} \epsilon_{j \lambda} \braket{\mathcal{O}^{J/\psi}(^3S_1^{[8]})} \, , \\
\label{eq: 3S18 LDME}
    \end{aligned}
\end{equation}
When interested in unpolarized $J/\psi$, we sum over $\lambda$ which yields a $\delta_{ij}$ factor; otherwise we can select a specific helicity. The calculation in this paper always includes only boosted polarization vectors: ${\Lambda^\mu}_j  \epsilon^j_{ \lambda}  \equiv \epsilon^\mu_\lambda$.

We can relate this representation of the $J/\psi$ polarization to the parametrization presented in Refs. \cite{Bacchetta:2000jk, Kumano:2020gfk}. The product of two polarization vectors for a given helicity can be broken down into unpolarized, vector polarized, and tensor polarized parts:
\begin{equation}
     \epsilon^*_{\lambda i} \epsilon_{j \lambda} = \frac13 \delta_{ij} + \frac{i}{2} \epsilon_{ijk}S_k - T_{ij} \, ,
\end{equation}
where 
\begin{equation}
\label{eq: S vec}
\begin{aligned}
    \vec{S} = {\rm Im}\left(\epsilon_\lambda^* \times \epsilon_\lambda\right) = (S^x_T, S_T^y, S_L)
\end{aligned}
\end{equation}
and 
\begin{equation}
\label{eq: Tij}
\begin{aligned}
    T_{ij} = & \; \frac13 \delta_{ij} - {\rm Re}\left( \epsilon^*_{\lambda i} \epsilon_{j \lambda}\right) \\
     \equiv & \; \frac12
\begin{pmatrix}
-\frac23 S_{LL} + S_{TT}^{xx} & S_{TT}^{xy} & S_{LT}^x\\
S_{TT}^{yx} & -\frac23 S_{LL} - S_{TT}^{xx} & S_{LT}^y\\
S_{LT}^x & S_{LT}^y & \frac43 S_{LL}
\end{pmatrix}.
\end{aligned}
\end{equation}
We refer readers to the appendixes of Ref. \cite{Bacchetta:2000jk} for physical interpretations of these tensor parameters. They take on specific values upon choice of a helicity $\lambda$. Applying the boost matrices gives
\begin{equation}
    {\Lambda^\mu}_i {\Lambda^\nu}_j \epsilon^{*}_i \epsilon_{j} = -\frac13 \left( g^{\mu \nu} -\frac{P^\mu P^\nu}{M^2} \right) + \frac{i}{2M} \epsilon^{\mu\nu\alpha\beta}P_\alpha S_\beta - T^{\mu\nu} \, ,
\label{eq: Jpsi Pol}
\end{equation}
where the boosted spin vector and tensor are 
\begin{equation}
    S_{\beta} = \left(P^+ \bar{n}_\beta - P^- n_\beta\right) \frac{S_L}{M} + S_{T\beta} \; ,
\label{eq: spin vector}
\end{equation}
and 
\begin{equation}
\begin{aligned}
    T^{\mu \nu} = \frac12\bigg\{\left[\frac43 \frac{(P^+)^2}{M^2} \bar{n}^\mu \bar{n}^\nu - \frac23 \bar{n}^{\{\mu}\bar{n}^{\nu\}} + \frac13\frac{M^2}{(P^+)^2} n^{\mu}n^{\nu}\right]S_{LL}\\
    -\frac{1}{M}(P^+ \bar{n} - \frac{M^2}{2P^+}n)^{\{\mu} S^{~\nu\}}_{LT} +\frac23 S_{LL}g^{\mu\nu}_T + S_{TT}^{\mu\nu}\bigg\} \; ,
\end{aligned}
\end{equation}
respectively. 

The polarization vectors describe either transverse or longitudinal polarizations with respect to the $J/\psi$'s motion. The transverse polarizations correspond to $\lambda$ = $\pm$ 1 and for the frames we work with they are explicitly given by
\begin{equation}
    \epsilon_{\pm} = \frac{1}{\sqrt{2}} (\mp1, -i, 0).
\end{equation}
Similarly, the longitudinal polarization is $\lambda$ = 0,
\begin{equation}
    \epsilon_{0} = (0, 0, 1).
\end{equation}

Plugging these vectors into Eqs. (\ref{eq: S vec}) and (\ref{eq: Tij}) allows one to easily solve for the values of the spin parameters $S_L$, $S_{LL}, S^{\mu}_T$, etc. In particular, for transverse polarizations we find all parameters vanish except for $S_L = \pm 1$ for the $\lambda = \pm 1$ states, respectively, and $S_{LL} = \frac12$ for both. Similarly for pure longitudinal polarization, we find only $S_{LL} = -1$ and all other parameters vanish. 
However, to identify the possible polarized TMDFFs for $J/\psi$, it is important to leave these parameters general, as they play a key role in the classification process. 

\subsection{Quark TMDFF projections}
\label{sec: quark pol}

\begin{table*}
\begin{tabular}{|c|c||c|c|c|}
 \cline{3-5} 
 \multicolumn{2}{c||}{\multirow{2}{*}{}} & \multicolumn{3}{c|}{Quark polarization} \\ \cline{3-5}
 \multicolumn{2}{c||}{} & Unpolarized & Longitudinal & Transverse \\ \hline \hline
 \multirow{6}{0.5cm}{\rotatebox[origin=c]{90}{ Hadron polarization }} & U & $D_1$ & & $H_1^\perp$ \\ \cline{2-5}
 & L &  & $G_1$ & $H_{1L}^\perp$ \\ \cline{2-5}
 & T & $D_{1T}^\perp $ & $G_{1T}^\perp$ & $H_1,\, H_{1T}^\perp$ \\ \cline{2-5} 
 & LL & $D_{1LL}$ & & $H_{1LL}^\perp$ \\ \cline{2-5}
 & LT & $D_{1LT}$ & $G_{1LT}$ & $H_{1LT}^\perp , \, H_{1LT}^\prime$ \\ \cline{2-5}
 & TT & $D_{1TT}$ & $G_{1TT}$ & $H_{1TT}^\perp, \, H_{1TT}^\prime $ \\ \hline
\end{tabular}
\caption{Possible leading quark TMDFFs arising from the combinations of unpolarized, longitudinal, or transversely polarized quarks and unpolarized, vector, or tensor polarizations of a spin-1 hadron. Note the labels L, T, LL, LT, and TT indicate the subscript of the spin parameter to which the fragmentation function is proportional. In particular, L and T polarizations do not represent longitudinally and transversely polarized $J/\psi$'s as they are conventionally understood. }
\label{tab: quark TMDFFs}
\end{table*}


The procedure for classifying polarized TMD PDFs and FFs was originally developed for spin-1/2 quarks and spin-1/2 external hadrons by Refs. \cite{Boer:1997nt,Tangerman:1994eh,Mulders:1995dh}. This yields eight independent nonperturbative TMDs. The process for identifying the possible polarized TMD PDFs and FFs for a spin-1/2 quark and a spin-1 external hadron is nearly identical. This was first performed by Bachetta and Mulders \cite{Bacchetta:2000jk} by writing down all tensor structures that are invariant under parity and hermiticity. This yields 18 possible TMDFFs which are summarized in Table \ref{tab: quark TMDFFs}. 

In practice, we project out the TMDFFs from Eq.~(\ref{eq: q TMDFF}) in the following way. First, we note the initial spin-1/2 quark can either be unpolarized, longitudinally polarized, or transversely polarized at leading twist. This corresponds to taking $\Gamma \to \gamma^+/2, ~\gamma^+\gamma_5/2,$ or $ i\sigma^{\beta +}\gamma_5/2$, respectively, and then evaluating the trace in Eq.~(\ref{eq: q TMDFF}). Then, the polarization vectors that appear with the LDME in Eq.~(\ref{eq: 3S18 LDME}) are boosted and parametrized according to Eq.~(\ref{eq: Jpsi Pol}). This produces many objects proportional to the same spin parameters characterized by Ref. \cite{Bacchetta:2000jk}. These tensor objects produced by this procedure are not the TMDFFs themselves, but rather proportional to the TMDFFs. Thus finally, to isolate the distributions, we invert the factors identified in Ref.~\cite{Bacchetta:2000jk} and systematically apply them to project out the final results. These steps are summarized by the operations given in Appendix \ref{app: Projection} and the exact TMDFFs are determined by inverting the factors in Eqs. (\ref{eq: U quark FFs})-(\ref{eq: T quark FFs}). 

Using the definitions of the fragmentation functions and the polarization machinery presented, we calculate the TMDFFs in the $\boldsymbol{p}_\perp \gg \Lambda$ regime.
%
%
In momentum space, the leading-order (LO) nonzero FFs for an unpolarized quark to fragment to $J/\psi$ are
\begin{equation}
    \begin{aligned}
        D_1(z, \kv_T ; \mu) = & \; \frac{2\alpha_s^2(\mu)}{9\pi N_c M^3 z} \frac{\kv_T^2 z^2(z^2-2z+2)+2M^2(z-1)^2}{[z^2 \kv_T^2+M^2(1-z)]^{2}} \braket{\mathcal{O}^{J/\psi}(^3S_1^{[8]})}\; , \\
        D_{1LL}(z, \kv_T ; \mu) =  & \; \frac{2\alpha_s^2(\mu)}{9\pi N_c M^3 z}\frac{\kv_T^2z^2(z^2-2z+2)-4 M^2(z-1)^2}{[z^2\kv_T^2+M^2(1-z)]^{2}}\braket{\mathcal{O}^{J/\psi}(^3S_1^{[8]})}\; , \\
        D_{1LT}(z, \kv_T ; \mu) = & \; \frac{2\alpha_s^2(\mu)}{3\pi N_c M } \frac{(2-z)(1-z) }{[z^2\kv_T^2+M^2(1-z)]^{2}} \braket{\mathcal{O}^{J/\psi}(^3S_1^{[8]})} \; , \\
        D_{1TT}(z, \kv_T ; \mu) = & \; \frac{2\alpha_s^2(\mu)}{3\pi N_c M} \frac{z(z-1)}{[z^2\kv_T^2+M^2(1-z)]^{2}} \braket{\mathcal{O}^{J/\psi}(^3S_1^{[8]})}\; .
    \end{aligned}
\end{equation}
The nonzero FFs for a longitudinally polarized quark to fragment to $J/\psi$ are:
\begin{equation}
    \begin{aligned}
        G_{1L}(z, \kv_T ; \mu) = & \; \frac{\alpha_s^2(\mu)}{3\pi N_c M^3} \frac{ \kv_T^2 z^2(2-z)}{[z^2\kv_T^2+M^2(1-z)]^{2}}\braket{\mathcal{O}^{J/\psi}(^3S_1^{[8]})} \; , \\
        G_{1T}^\perp(z, \kv_T ; \mu) = & \;  \frac{2\alpha_s^2(\mu)}{3\pi N_c M} \frac{z(z-1)}{[z^2\kv_T^2+M^2(1-z)]^{2}} \braket{\mathcal{O}^{J/\psi}(^3S_1^{[8]})}\, . \\
    \end{aligned}
\end{equation}
All of the FFs for the transversely polarized quark, denoted by $H$,  as well as $D_{1T}^\perp$,  $G_{1LT}$, and $G_{1TT}$ vanish at LO in this calculation. The 
 transversely polarized quark FFs all vanish because when $\Gamma= i\sigma^{\beta+}\gamma_5/2$ is inserted into the expressions in Eqs. (\ref{DA})-(\ref{DC}) 
 the traces are over an odd number of $\gamma$ matrices.
Note that our answer for $D_1$ differs from Ref. \cite{Echevarria:2020qjk} by a factor of 1/3 because we choose to identify $D_1$ as the fragmentation function that comes from inserting the polarization projection $\frac13(-g^{\mu \nu} + \frac{P^\mu P^\nu}{M^2})$. This is for convenience when using $D_1$ in polarized cross sections. We have checked that our $D_1$ agrees with Ref. \cite{Echevarria:2020qjk} when we sum over all polarizations (effectively multiplying $D_1$ by 3).

\section{Gluon Fragmentation}
\label{sec: g fragmentation}

\begin{figure*}
    \includegraphics[trim=5.2cm 21.0cm 9.1cm 4.4cm,clip,scale=1]{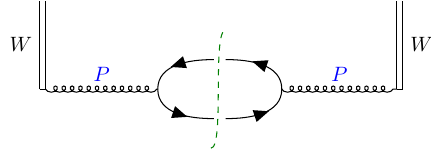}
    \caption{Tree-level diagram contributing to the gluon TMDFF.  Momenta are labeled in blue.}
    \label{gluondiag}
\end{figure*}

For completeness we also calculate the gluon TMDFF for fragmentation to $J/\psi$.  The procedure is essentially the same. Again, we employ the TMD NRQCD factorization theorem given by Eq.~(\ref{eq: NRQCD fact}). Starting from the definition in Eq.~(\ref{eq: g TMDFF}), we can calculate the short distance coefficients in perturbative QCD which can then be matched onto the long-distance matrix elements in NRQCD. At leading order in $\alpha_s$ there is only one possible diagram, shown in Fig. \ref{gluondiag}. 

The amplitude corresponding to the diagram in the figure is
\begin{equation}
\begin{aligned}
d^{\alpha \alpha'}_g = \frac{g^2}{2z^2P^+(N_c^2-1)}\int & \frac{db^-}{(2\pi)} e^{i(b^-P^+/z - P\cdot b)}\left(P^\alpha \frac{n^\mu}{P^2} - P^+ \frac{g^{\alpha\mu}}{P^2}\right)\chi_{\mu \nu}\\
&\times\left(P^{\alpha'} \frac{n^\nu}{P^2} - P^+ \frac{g^{\alpha'\nu}}{P^2}\right).
\label{eq: gTMDFF amp}
\end{aligned}
\end{equation}
Since the impact parameter is taken to have $b^+ = 0$, the integral over $b^-$ produces a $\delta$ function $1-z$ and Fourier transforming to $\kv_T$ space produces one in $\kv_T$,
\begin{equation}
    \int \frac{db^-}{(2\pi)} \frac{d^2{\bf b}_T}{(2\pi)^2} e^{-i(\kv_T \cdot {\bf b_T})} e^{i(b^-P^+ /z - b^-P^+ + \Pv_T \cdot {\bf b}_T)} = \frac{(z^2)_{z=1}}{P^+}\delta(1-z) \delta^{(2)}(\kv_T).
\end{equation}
The $\delta^{(2)}(\kv_T)$ arises from the fact that the $J/\psi$ has no transverse momentum in the frame we are working in ($\Pv_T = 0$) and there is no other final state particle at this order. This simplifies Eq.~(\ref{eq: gTMDFF amp}) to
\begin{equation}
\begin{aligned}
d^{\alpha \alpha'}_g = \frac{g^2}{2(P^+ M^2)^2(N_c^2-1)}\delta(1-z) \delta^{(2)}(\kv_T)\big(P^\alpha n_\mu - P_+ g^{\alpha}_{ \mu}\big)\chi^{\mu \nu}\big(P^{\alpha'} n_\nu - P_+ g^{\alpha'}_{\nu}\big).
%
\label{eq: gTMDFF amp 2}
\end{aligned}
\end{equation}
Notice the same spinor factor $\chi^{\mu\nu}$ has appeared in this calculation so the matching procedure is exactly the same as for the case of a quark fragmenting to a $J/\psi$. The nonrelativistic expansion in Eq.~(\ref{eq: NR spinors}) is performed again and then matched onto the long-distance matrix element, Eq.~(\ref{eq: 3S18 LDME}).

\subsection{Gluon TMDFF projections}
\label{sec: gluon pol.}

The gluon is a spin-1 particle and hence there is a different procedure for identifying the possible processes when it fragments into a spin-one hadron. However, the ideas are similar to the decomposition made for the $J/\psi$ polarization. The gluon polarization vectors can be decomposed into their scalar, antisymmetric, and symmetric traceless components corresponding to unpolarized, helicity-0, and helicity-2 gluon states.

This decomposition was first made by Boer \emph{et.~al.}~\cite{Boer:2016xqr} for TMD PDFs of spin-1 gluons inside of spin-1 hadrons. The process for identifying the TMDFFs for spin-1 gluons fragmenting to spin-1 hadrons is the same if one make the replacements \{$n, P^+, x$\} $\to$ \{$\bar{n}, P^-, 1/z$\} . To project out these fragmentation functions from the definition in Eq.~(\ref{eq: gTMDFF amp}) we again insert the parametrization for the polarization vectors given in Eq.~(\ref{eq: Jpsi Pol}). Then, similar to the quark case, we invert the expressions given in Ref. \cite{Boer:2016xqr}. The relevant pieces are listed in Appendix \ref{app: Projection} for convenience. 

\begin{table*}
\begin{tabular}{|c|c||c|c|c|}
 \cline{3-5}
 \multicolumn{2}{c||}{\multirow{2}{*}{}} & \multicolumn{3}{c|}{Gluon operator polarization} \\ \cline{3-5}
 \multicolumn{2}{c||}{} & Unpolarized & Helicity 0 antisymmetric & Helicity 2 \\ \hline \hline
 \multirow{6}{0.5cm}{\rotatebox[origin=c]{90}{ Hadron polarization }} & U & $D_1^g$ & & $H_1^{\perp g}$ \\ \cline{2-5}
 & L &  & $G_{1L}^g$ & $H_{1L}^{\perp g}$ \\ \cline{2-5}
 & T & $D_{1T}^{\perp g} $ & $G_{1T}^{\perp g}$ & $H_{1T}^g,\, H_{1T}^{\perp g}$ \\ \cline{2-5}
 & LL & $D_{1LL}^g $ & & $H_{1LL}^{\perp g}$ \\ \cline{2-5}
 & LT & $D_{1LT}^g $ & $G_{1LT}^g$ & $H_{1LT}^{\perp g} , \, H_{1LT}^{\prime g}$ \\ \cline{2-5}
 & TT & $D_{1TT}^g $ & $G_{1TT}^g$ & $H_{1TT}^{\perp g}, \, H_{1TT}^{\prime g}$ \\ \hline
\end{tabular}
\caption{\label{tab:gluonTMDFFs}Leading gluon TMDFFs arising from the combinations of unpolarized, antisymmetric, or symmetric combinations of gluons and unpolarized, vector, or tensor polarizations of a spin-1 hadron. The hadron labels are the same as in Table \ref{tab: quark TMDFFs}.}
\end{table*}

From the general decomposition there are 18 gluon TMDFFs which we list in Table \ref{tab:gluonTMDFFs}. However at leading order in $\alpha_s$ and the relative velocity we find that almost all of these functions vanish. The nonzero TMDFFs at this order in the $\boldsymbol{p}_\perp \gg \Lambda$ region are 
\begin{equation}
    \begin{aligned}
        D^g_1(z, \kv_T ; \mu) = & \; \frac{\pi \alpha_s(\mu)}{9M^3}\braket{\mathcal{O}^{J/\psi}(^3S_1^{[8]})}\delta(1-z) \delta^{(2)}(\kv_T)\;  , \\
        D^g_{1LL}(z, \kv_T ; \mu)  = & \; \frac{\pi \alpha_s(\mu)}{9M^3}\braket{\mathcal{O}^{J/\psi}(^3S_1^{[8]})}\delta(1-z) \delta^{(2)}(\kv_T) \; , \\
        G^g_{1L}(z, \kv_T ; \mu)  =  & \; -\frac{\pi \alpha_s(\mu)}{6M^3}\braket{\mathcal{O}^{J/\psi}(^3S_1^{[8]})}\delta(1-z) \delta^{(2)}(\kv_T) \; , \\
        H^g_{1TT}(z, \kv_T ; \mu)  = & \; -\frac{\pi \alpha_s(\mu)}{6M^3}\braket{\mathcal{O}^{J/\psi}(^3S_1^{[8]})}\delta(1-z) \delta^{(2)}(\kv_T) \; .\\
    \end{aligned}
\end{equation}
Notice again that the gluon fragmentation functions are only proportional to $\delta^{(2)}(\kv_T)$and have no other $\kv_T$ dependence. Again, $D^g_1$ differs from the conventional unpolarized fragmentation function \cite{Echevarria:2023dme} by a factor of 1/3 because of the projection $\frac13(-g^{\mu \nu} + \frac{P^\mu P^\nu}{M^2})$.

%

At the next order in $\alpha_s$ the picture quickly becomes more complicated. Since real gluon emission is permitted, in addition to the exchange of virtual gluons or quarks, 
this will introduce more sophisticated features into the calculation, such as nontrivial $\kv_T$ dependence, rapidity divergences and mixing between the LDMEs. 
To access the transverse momentum dependence of various structure functions via gluon fragmentation to $J/\psi$ it will be crucial to understand these higher-order contributions as well. These corrections were recently calculated for $D^g_1$ for the first time in Ref. \cite{Echevarria:2023dme}. We will present a similar analysis for the other polarized TMDFFs in future work.

\section{Phenomenology} 
\label{sec: phenom}

The fragmentation functions computed in this work are universal functions and can, in principle, be applied to a wide range of physical processes provided factorization theorems for these processes exist. Possible examples include $J/\psi$ production in jets, $e^+e^-$ annihilation, and SIDIS. The extraction of $J/\psi$ TMDFFs from $e^+e^-$ annihilation at small $P_T$ offers a clean way to verify our results experimentally \cite{Collins:1981uk, Boer:2008fr, Pitonyak:2013dsu}. On the other hand, the application of our results to SIDIS promises to give an alternative method to  access to the quark TMD PDFs in the proton \cite{Bacchetta:2006tn, Bacchetta:2000jk}.  In this section we write down and plot cross sections for $J/\psi$ production via SIDIS which are sensitive to the quark FFs, by analogy with Ref.~\cite{Echevarria:2020qjk}.

\subsection{SIDIS}
At leading twist, the TMDFFs for quarks fragmenting to a $J/\psi$ appear in the factorization theorems for SIDIS, which is highly relevant for the upcoming EIC \cite{AbdulKhalek:2021gbh,Accardi:2012qut}. SIDIS is the reaction
\begin{equation}
   \ell(l) + h(p) \rightarrow \ell(l^\prime)+H(P) + X \, ,
\end{equation}
where $\ell$ is a lepton, $h$ is the initial nucleon, $H$ is the final state $J/\psi$, and $X$ is the undetected part of the final state. 
\begin{figure*}
    \includegraphics[trim=5.2cm 18.8cm 12cm 4.3cm,clip,scale=1.2]{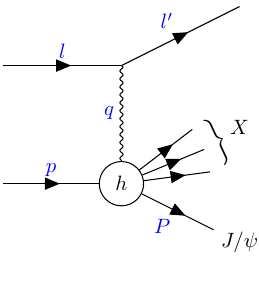}
    \caption{Leading order in $\alpha$ diagram for semi-inclusive deep inelastic scattering.  Momenta are labeled in blue.}
\end{figure*}
The differential cross section for SIDIS is given by \cite{Bacchetta:2000jk,Bacchetta:2006tn,Echevarria:2020qjk}:
\begin{equation}
   \frac{d\sigma}{dx \, dz \, dQ^2 \, d{\bf P}_\perp^2} = \frac{\alpha_{\rm em}^2 M}{2Q^2xzs} L^{\mu\nu}W_{\mu\nu} \, ,
\end{equation}
where we define the usual kinematic variables \\
\begin{equation}
    Q^2 = -q^2 = -(l-l')^2, ~~x = \frac{Q^2}{2p\cdot q}, ~~y= \frac{p\cdot q}{p\cdot l}, ~~z = \frac{p\cdot P}{p\cdot q}.
\end{equation}\\
The leptonic and hadronic tensors are
\begin{equation}
    L_{\mu \nu} = e^{-2} \bra{l'} J_\mu(0)\ket{l}\bra{l} J^\dagger_\nu (0) \ket{l'} \, .
\end{equation}
\begin{equation}
W_{\mu \nu} = e^{-2}\int \frac{d^4 x} {(2\pi)^4} e^{-ixq} \sum_X \bra{p} J_\mu^\dagger(x) \ket{P,X}\bra{P,X} J_\nu(0) \ket{p} 
\end{equation}
For production via fragmentation, TMD factorization proceeds in a similar manner to usual SIDIS factorization \cite{Collins:2011zzd,Echevarria:2011epo,Echevarria:2012js,Chiu:2012ir}, and allows the hadronic tensor to be written at leading twist as \cite{Bacchetta:2000jk}
\begin{equation}
    W^{\mu \nu} = 2 z \int d^2\kv_T \, d^2\pv_T ~\delta^{(2)}\left(\pv_T - \kv_T + \frac{\Pv_\perp}{z} \right){\rm Tr} \left[\gamma^\mu \Phi(\pv_T, x) \gamma^{\nu} \Delta(\kv_T, z) \right] \, 
\label{eq: Factorized W}
\end{equation}
where $\Phi(\pv_T, x)$ is a function of the possible TMD PDFs in the proton and $\Delta(\kv_T, z)$ describes the possible fragmentation functions. Here, the final transverse momentum of the $J/\psi$, $\Pv_\perp$, is proportional to the difference between the fragmenting quark's transverse momentum, $\kv_T$ and the initial parton's transverse momentum $\pv_T$.

The SIDIS cross sections are presented for quark fragmentation in Refs.~\cite{Bacchetta:2000jk, Bacchetta:2006tn}. Since many of the convolutions in these expressions vanish, either due to the vanishing fragmentation functions at leading order or from evaluating the convolution integrals in the structure functions, we only need to consider a few contributions. More precisely, we find all contributions with a transversely polarized light quark vanish at leading order. Additionally, all convolution integrals which contain odd factors of $k_x$, $k_y$, or $k_x^2 - k_y^2$ vanish because the fragmentation functions and parton distribution functions are even in $\kv_T^2$ at leading order. The nonvanishing contributions to the cross sections are presented in Appendix \ref{app: cross sections}, where the different structure functions from polarized $J/\psi$ are included as well.

\subsection{Numerical analysis}
\label{sec: results}

Using the factorized cross sections defined in Appendix \ref{app: cross sections}, we can make predictions.
%
For the SIDIS cross sections we also need the parton TMD PDFs. These quantities are still poorly constrained and it is an ongoing effort to determine their exact transverse momentum dependence \cite{Boussarie:2023izj, Musch:2007ya, Anselmino:2013lza, Sun:2014dqm, Scimemi:2019cmh, Bertone:2019nxa}. A simple parametrization of TMD PDFs often used in the literature is
\begin{equation}
    \Phi_{i/N}(x, \pv_T) = \frac{1}{\pi \left<p^2_T\right>}\Phi_{i/N}(x)e^{-{\pv_T^2}/\braket{p^2_T}}
\end{equation}
%
where $\Phi_{i/N}(x)$ represents the corresponding collinear PDF. The parameter $\left<p^2_T\right>$ varies for each polarized TMD PDF and parton that is considered. However, we find that, after varying $\left<p^2_T\right>$ between 0.2 and 0.8 GeV$^2$, the Gaussian approximation is not sufficiently different when applied in Eq.~(\ref{eq: Factorized W}) from the first-order approximation in the $\pv_T \gg \Lambda$ TMD expansion (in which the transverse momentum dependence is a $\delta$ function in $\pv_T$). Thus, for the observables considered in this paper we find it efficient to use

%
%
%
\begin{equation}
\begin{aligned}
    \Phi_{i/N}(x,\pv_T) &\approx \Phi_{i/N}(x) \delta^{(2)}(\pv_T)\\
    \end{aligned}
\end{equation}
%


\begin{figure}
    \centering
    \includegraphics[width =.9\linewidth]{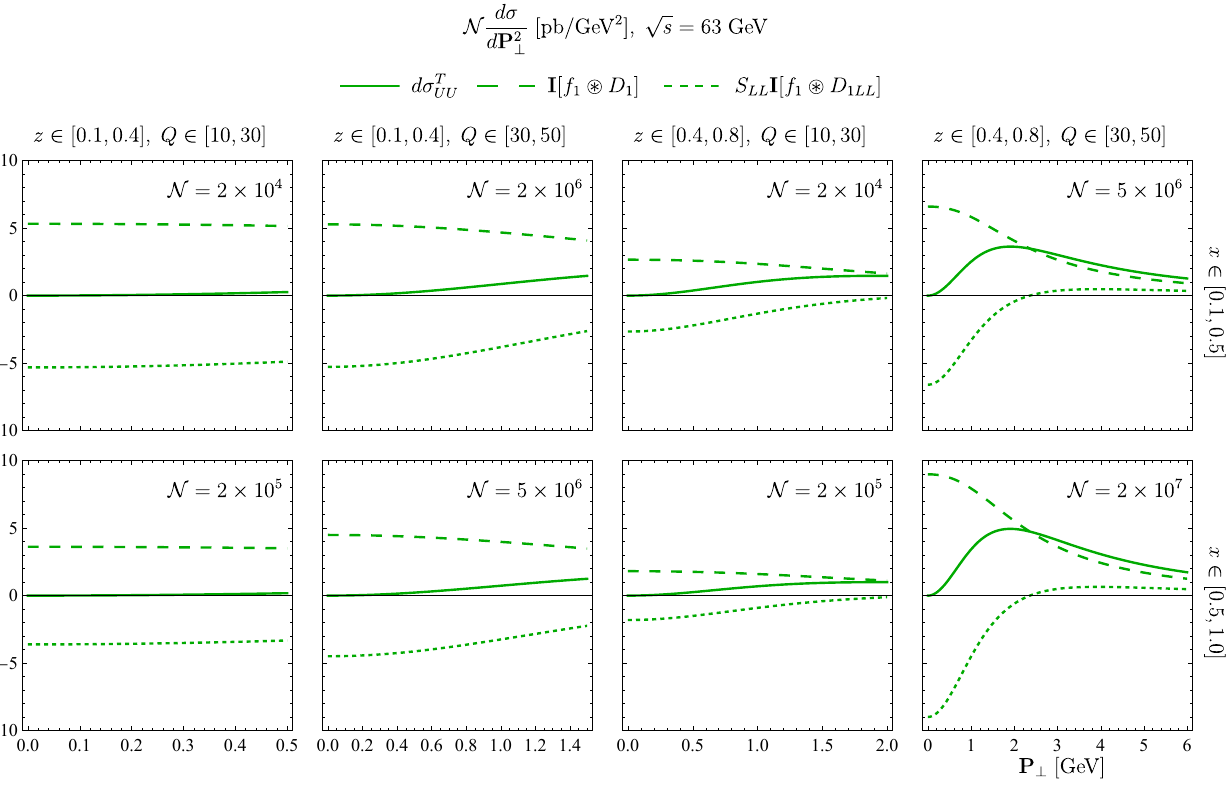}
    \caption{Production of transversely polarized $J/\psi$ from an unpolarized beam colliding with an unpolarized target in SIDIS. Solid green line shows the total cross section, while the dashed line gives the contribution from unpolarized fragmentation and the dotted line gives the contribution from the $D_{1LL}$ fragmentation function with $S_{LL} = 1/2$.}
    \label{fig: T cross sections}
\end{figure}
For SIDIS with an unpolarized lepton beam and an unpolarized target, the production of $J/\psi$ is given by Eq.~(\ref{eq: UU fac cross}). In the matching of the TMDFF in Eq.~(\ref{eq:matchingNRQCD}), $J/\psi$ production occurs at a scale roughly around the mass of the $J/\psi$ so we evaluate the strong coupling at $\mu = 3.1$ GeV in the $J/\psi$ fragmentation functions. However the PDF is probed at the scale $\sim Q$, so we choose $\mu = 30$ GeV for the PDFs. In the numerical analysis we use the PDF sets for the up and down quarks from Ref.~\cite{Bastami:2018xqd}.

One of the purposes of this work is to provide a more solid theoretical framework in the small $\Pv_\perp$ region so that the NRQCD LDMEs of $J/\psi$ can be properly extracted from a global analysis of the world's data. We leave this for a future study and choose the values determined by Chao {\it et.~al.~}\cite{Chao:2012iv}, setting $\braket{\mathcal{O}^{J/\psi}(^3S_1^{[8]})}$ = $0.3 \times 10^{-2}$ GeV in this analysis. 


To compare with the upcoming EIC, we study the cross section at the center of mass energy $\sqrt{s} = 63$ GeV. Following Ref. \cite{Echevarria:2020qjk} we divide the kinematic phase space into several bins of $x \in [0.1, 0.5]$ and $[0.5,1]$, $z \in [0.1, 0.4]$ and $[0.4,0.8]$, and $Q $(GeV) $\in [10, 30]$ and $[30, 50]$. This yields eight possible regions of interest. In each region, we plot in the TMD regime $\Pv_\perp \in [0, z_{\rm [bin~ min]} Q_{\rm [bin ~min]}/2]$ \cite{Echevarria:2020qjk}.


%
\begin{figure}
    \centering
    \includegraphics[width = .9\linewidth]{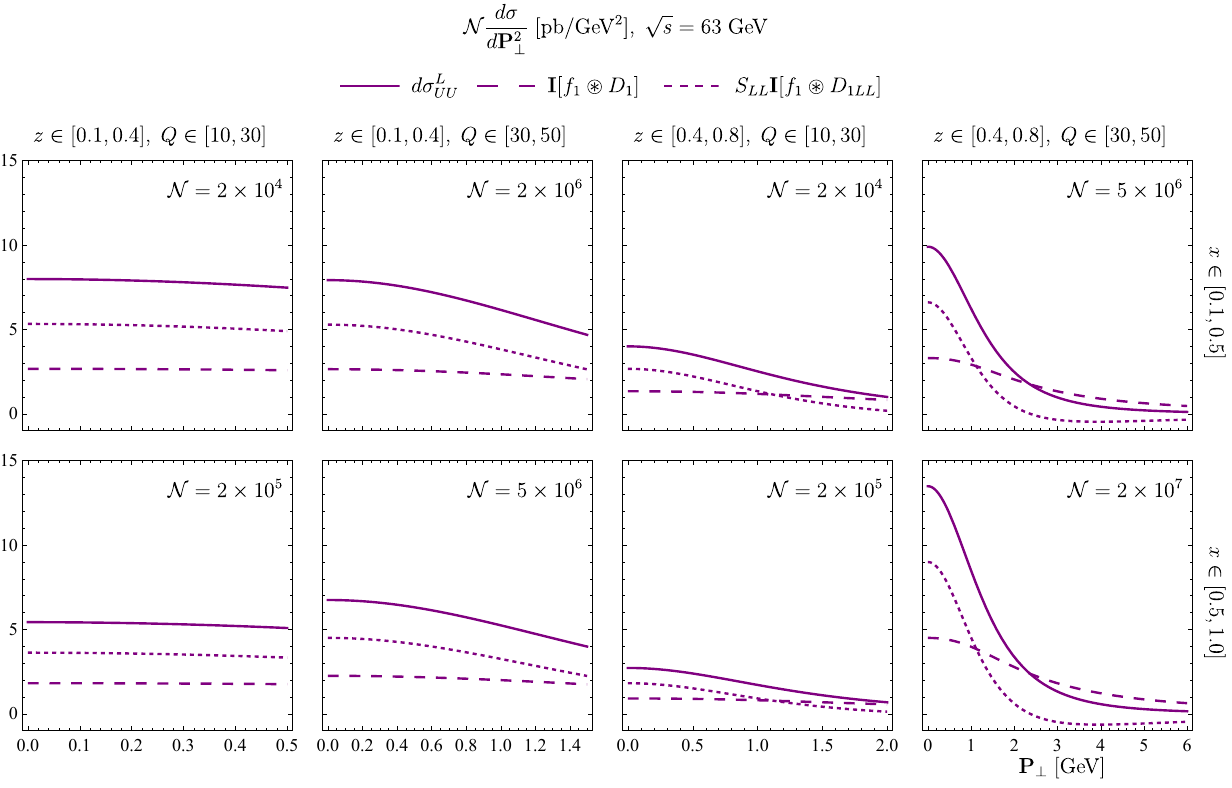}
    \caption{Production of longitudinally polarized $J/\psi$ from an unpolarized beam colliding with an unpolarized target in SIDIS. Solid purple line shows the total cross section, while the dashed line gives the contribution from unpolarized fragmentation and the dotted line gives the contribution from the $D_{1LL}$ fragmentation function (with $S_{LL} = -1$).}
    \label{fig: L cross sections }
\end{figure}

\begin{figure}
    \centering
    \includegraphics[width = .9\linewidth]{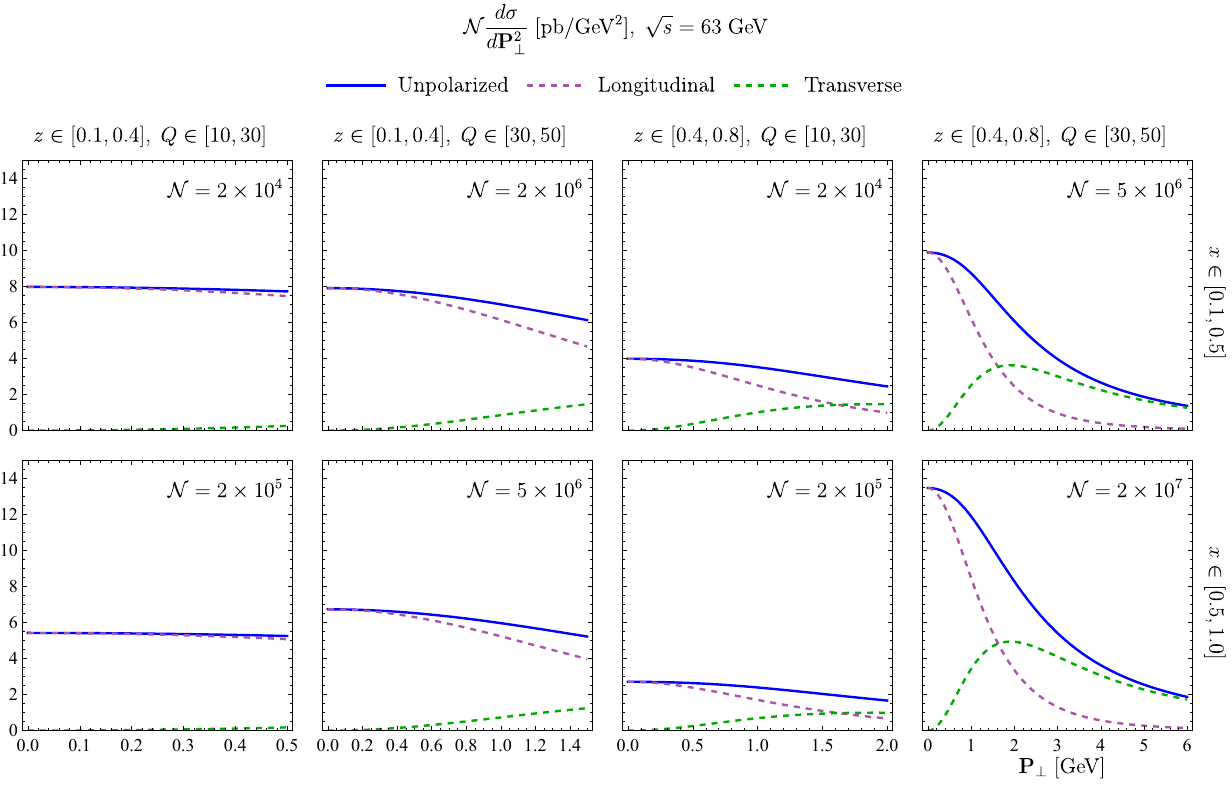}
    \caption{Production of unpolarized (blue, solid), longitudinally polarized (purple, dashed), and transversely polarized (green, dashed) $J/\psi$ from an unpolarized beam colliding with an unpolarized target in SIDIS. }
    \label{fig: A cross sections }
\end{figure}

%
The decomposition of the polarization vectors in Eq.~(\ref{eq: Jpsi Pol}) allows us to study the production of transversely polarized and longitudinally polarized $J/\psi$ in this process. As discussed in Sec. \ref{sec: Jpsi pol}, for transversely polarized $J/\psi$, the spin parameters should be set to $S_{LL}$ = 1/2 and $S_{L} = \pm 1$ (for $\lambda = \pm 1$, respectively) with all others being zero. Of course the factorized cross section in Eq.~(\ref{eq: UU fac cross}) is independent of $S_L$, so the latter point is not relevant here. Since a $\lambda= +1$ polarization is indistinguishable from $\lambda = -1$, the two should be added together to compare with experiment. Our predictions for the cross section of transversely polarized $J/\psi$ are presented in Fig. \ref{fig: T cross sections}. Qualitatively, we observe that the structure function proportional to $S_{LL}$ suppresses the production of transverse $J/\psi$, especially at smaller ranges of $\Pv_\perp$. To this extent, the cross section for transverse $J/\psi$ is essentially zero around $\Pv_\perp \approx 0$, regardless of the kinematic region.

Similarly, we can plot the production of longitudinally polarized $J/\psi$. This is presented in Fig. \ref{fig: L cross sections }. Here, the correct parametrization is to set $S_{LL}=-1$ and all other spin parameters to zero. In this case, the change in sign of $S_{LL}$ causes the structure function to enhance the production of longitudinal $J/\psi$. Now note that none of the cross sections are zero around $\Pv_\perp \approx 0$.

If we sum over all polarizations by adding the transverse and longitudinal $J/\psi$ cross sections together we can study the production of unpolarized $J/\psi$. This has the effect of canceling out the structure function dependent on $D_{1LL}$ in $d\sigma_{UU}$. In Fig. \ref{fig: A cross sections } we plot the unpolarized $J/\psi$ cross section, as well as the total transverse and longitudinal $J/\psi$ cross sections from Figs. \ref{fig: T cross sections} and \ref{fig: L cross sections }. The main observation is that the $J/\psi$ is predominantly longitudinal at low $\Pv_\perp$ and becomes more transverse at larger $\Pv_\perp$.

\begin{figure}
    \centering
    \includegraphics[width = .9\linewidth]{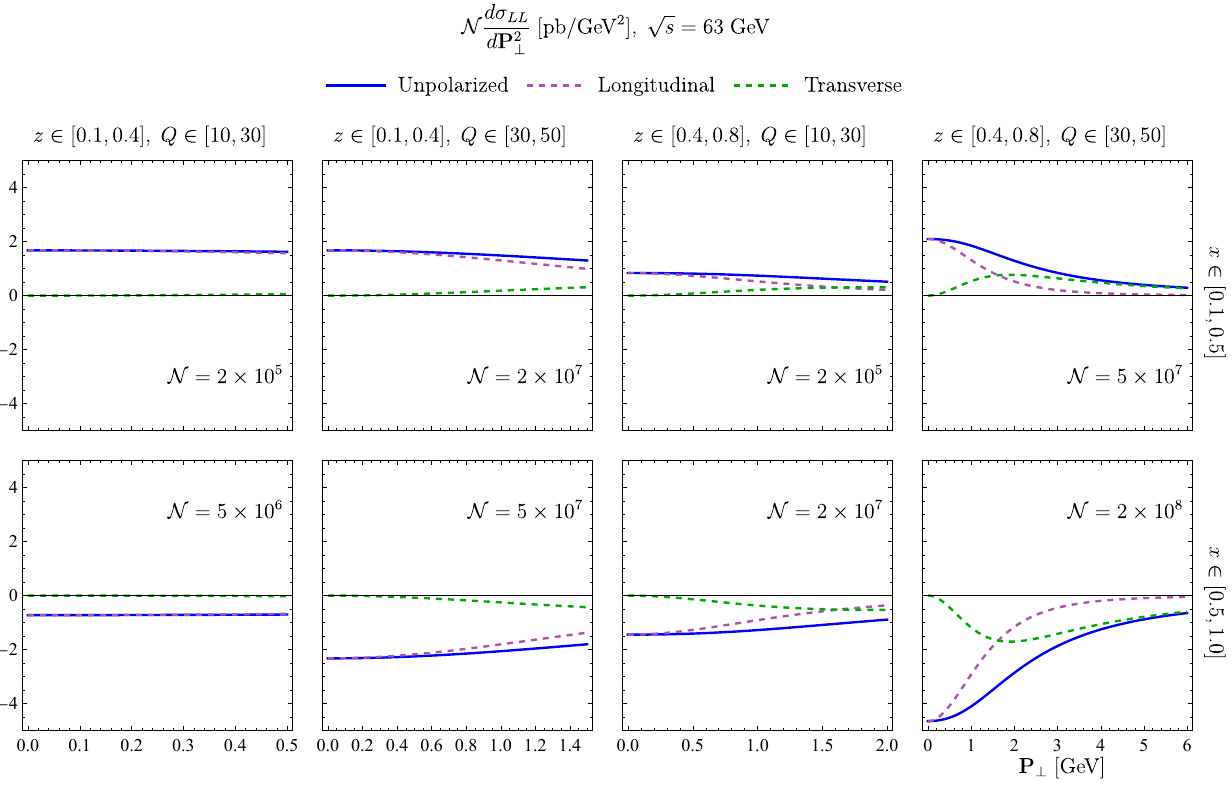}
    \caption{Production of unpolarized (blue, solid), longitudinally polarized (purple, dashed), and transversely polarized (green, dashed) $J/\psi$ from a longitudinal polarized beam colliding with a longitudinal polarized target in SIDIS. }
    \label{fig: A cross sections LL }
\end{figure}

It is also interesting to study SIDIS with polarized beams and targets. The observable for a longitudinally polarized beam and a longitudinally polarized target, is $d\sigma_{LL}$  which is defined as a difference between physical cross sections. In a helicity basis where superscripts represent the helicities of a nucleon target and subscripts represent the helicities of the virtual photon, $d\sigma_{LL}$ is \cite{Bacchetta:2006tn}
\begin{equation}
    d\sigma_{LL} = \frac12(d\sigma^{++}_{++} - d\sigma^{--}_{++}).
\end{equation}
The factorized expression for $d\sigma_{LL}$ is given in Eq.~(\ref{eq: LL cross section}). For the beam helicity we use $\lambda_e=-1$ which corresponds to a purely left-handed beam and for the quark polarization we choose $S_{qL} = -1$ putting the target spin parallel to the photon momentum \cite{Bacchetta:2006tn, Diehl:2005pc}. Notice this cross section is dependent on the polarized quark PDF, $g_{1L}(x, \pv_T)$. This quantity is poorly constrained. Even the collinear PDF $g_{1L}(x)$ has yet to be extracted precisely at the scales we are working. Nonetheless, the PDF can be evolved to $\mu = 30$ GeV and we present preliminary results. 

The cross sections for longitudinal beams scattering off of longitudinal targets are plotted in Fig. \ref{fig: A cross sections LL }. The striking observation is that, in the larger $x$ bin (bottom row), all of the curves are negative. Quantitatively, this can attributed entirely to the fact that the sum of the $g_{1L}(x)$ PDFs over flavors goes negative in this region. Qualitatively, this is not necessarily cause for concern. Since $d\sigma_{LL}$ is not a physical cross section, but rather a difference of cross sections, it is reasonable for it to turn negative and indicates that in this range of $x$, the negative helicity photon cross section dominates.

%
\section{Conclusion}

The main result of this paper is the calculation, at leading order in $\alpha_s(2m_c)$, of the matching coefficient of polarized TMDFFs onto NRQCD TMDFFs for the fragmentation of light quarks and gluons into $J/\psi$.  There are 18 possible quark TMDFFs, and we find that all but six of these have vanishing matching coefficient at LO. There are also 18 possible gluon TMDFFs and we find that all but four vanish. 

These results can be applied directly to a process of interest for the future EIC: SIDIS production of $J/\psi$. As an example, we use our results to calculate the fragmentation contribution to polarized $J/\psi$ production from both unpolarized beams and targets and from polarized beams and targets in SIDIS. We find in both cases that, for the bins considered, the production of longitudinally polarized $J/\psi$ dominates. 

Fragmentation of partons is not the only important mechanism for $J/\psi$ production in SIDIS. In addition to the quark fragmentation contribution considered in this paper, polarized $J/\psi$ production in SIDIS can occur through direct production in photon-gluon fusion. This process has been studied before in both collinear and TMD frameworks in Refs.~\cite{Fleming:1997fq, Bacchetta:2018ivt, Boer:2020bbd, DAlesio:2021yws,Boer:2023zit}. However, since quark fragmentation is comparable to photon-gluon fusion in many kinematic regimes \cite{Echevarria:2020qjk}, it is important that both mechanisms be included in a thorough analysis of polarized $J/\psi$ production in SIDIS.  
In addition, as we mentioned in Sec.~\ref{sec: results}, a rigorous framework for $J/\psi$ production in SIDIS will allow SIDIS data to be included in the  extraction of the NRQCD long-distance matrix elements from an analysis of the world's data. The $\braket{{\cal O}(^3S_1^{[8]})}$ is particularly difficult to identify, as several studies have given conflicting values \cite{Butenschoen:2011yh,Butenschoen:2012qr, Bodwin:2014gia, Chao:2012iv}.
In the kinematic regime $\boldsymbol{p}_\perp \sim \Lambda$, higher-order corrections to the NRQCD TMDFF will become necessary. This will introduce nontrivial $\Pv_\perp$ dependence into the TMDFF that could arise from soft gluon emission or other nonperturbative effects during the $c\bar{c}$ hadronization into a $J/\psi$. All of these effects  will be studied carefully in future work.  

{\bf Acknowledgments} - We thank Alexey Prokudin for helpful discussions and for giving access to the JAM Collaboration TMD PDFs. M.~C., R.~H., and T.~M. are supported by the U.S. Department of Energy, Office of Science, Office of Nuclear Physics under Award No.~DE-FG02-05ER41367. R.~H. and T.~M. are also supported by  the Topical Collaboration in Nuclear Theory on Heavy-Flavor Theory (HEFTY) for QCD Matter under Grant No.~DE-SC0023547. M.C. is supported by the National Science Foundation Graduate Research Fellowship under Grant No.~DGE 2139754. R.~G. and S.~F. are supported by the U.S. Department of Energy, Office of Science, Office of Nuclear Physics,  under Award No.~DE-FG02-04ER41338. 
\appendix

\section{Factorized SIDIS cross sections}
\label{app: cross sections}

Evaluating the hadronic tensor trace in Eq.~(\ref{eq: Factorized W}) yields many contributions to the cross section. The initial quark parton can be unpolarized, longitudinally polarized, or transversely polarized and the $J/\psi$ can be unpolarized, vector polarized, or tensor polarized as described in Sec.~\ref{sec: quark pol}. Many of the cross sections are zero for our purposes, as explained in Sec.~\ref{sec: phenom}. Here we list the nontrivial cross sections that we consider. More examples of cross sections for production of polarized spin-1 particles by polarized partons can be found in Refs. \cite{Bacchetta:2000jk, Bacchetta:2006tn}.

The nonvanishing leading order cross section for an unpolarized lepton to scatter off an unpolarized target is
\begin{equation}
\begin{aligned}
\label{eq: UU fac cross}
    \frac{d \sigma_{UU}(l + H \to l' + J/\psi + X)}{dx ~dz ~dy ~d^2 {\bf P_\perp}} 
    = &\frac{4\pi \alpha^2 s}{Q^4} \left(1 - y +\frac{y^2}{2}\right) \bigg\{{\bf I}[f_1 D_1] + S_{LL} {\bf  I}[f_1 D_{1LL}]\bigg\}\\
\end{aligned}
\end{equation}
We also consider a polarized lepton beam and a longitudinally polarized target. The leading order nonvanishing result is
\begin{equation}
    \frac{d \sigma_{LL}(l + H \to l' + J/\psi + X)}{dx ~dz ~dy ~d^2 {\bf P_\perp}} = \frac{4\pi \alpha^2 s}{Q^4}2 \lambda_c S_{qL} ~ y\bigg(1- \frac{y}{2}\bigg)x\bigg\{{\bf I}[g_{1L}D_1] + S_{LL} {\bf I}[g_{1L}D_{1LL}]\bigg\}.
\label{eq: LL cross section}
\end{equation}
where the convolution integral is defined as 
\begin{equation}
    {\bf  I}[f~ D] = 2 z \int d^2\kv_T \, d^2\pv_T ~\delta^{(2)}\left(\pv_T - \kv_T + \frac{\Pv_\perp}{z} \right) f(\pv_T) D(\kv_T).
\end{equation}

\section{Projection operators for the TMDFFs}
\label{app: Projection}

In this appendix, we summarize the operators that project out the individual quark and gluon TMDFFs \cite{Bacchetta:2000jk,Boer:2016xqr}. Here $\Delta_{\rm pol}^{[\Gamma]}$ is the quark FF proportional to the $J/\psi$ polarization parameter $S_{\rm pol}$ with quark polarization projection operator $\Gamma$,
\begin{equation}
    \begin{aligned}
        \Delta_U^{[\gamma^+]}(x,\kv_T) = & \; D_1 \; , \\
        \Delta_L^{[\gamma^+]}(x,\kv_T) = & \; 0 \; , \\
        \Delta_T^{[\gamma^+]}(x,\kv_T) = & \; \frac{1}{M}\epsilon_T^{\mu\nu}S_{T\; \nu} k_{T\;\mu} D_{1T}^\perp \; , \\
        \Delta_{LL}^{[\gamma^+]}(x,\kv_T) = & \; S_{LL} D_{1LL} \; , \\
        \Delta_{LT}^{[\gamma^+]}(x,\kv_T) = & \; \frac{1}{M}{\bf S}_{LT} \cdot \kv_T D_{1LT} \; , \\
        \Delta_{TT}^{[\gamma^+]}(x,\kv_T) = & \; \frac{1}{M^2} \kv_T \cdot {\bf S}_{TT} \cdot \kv_T D_{1TT} \; ,
    \end{aligned}
\label{eq: U quark FFs}
\end{equation}
\begin{equation}
    \begin{aligned}
        \Delta_U^{[\gamma^+ \gamma_5]}(x,\kv_T) = & \; 0 \; , \\
        \Delta_L^{[\gamma^+ \gamma_5]}(x,\kv_T) = & \; S_L G_{1L} \; , \\
        \Delta_T^{[\gamma^+ \gamma_5]}(x,\kv_T) = & \; \frac{1}{M} {\bf S}_T \cdot \kv_T G_{1T} \; , \\
        \Delta_{LL}^{[\gamma^+ \gamma_5]}(x,\kv_T) = & \;  0 \; , \\
        \Delta_{LT}^{[\gamma^+ \gamma_5]}(x,\kv_T) = & \; \frac{1}{M}\epsilon_T^{\mu \nu} S_{LT \; \nu} k_{T \; \mu} G_{1LT} \; , \\
        \Delta_{TT}^{[\gamma^+ \gamma_5]}(x,\kv_T) = & -\frac{1}{M^2}\epsilon_T^{\mu\nu} S_{TT \; \nu \rho} k_T^\rho k_{T\; \mu} G_{1TT} \; ,
    \end{aligned}
\label{eq: L quark FFs}
\end{equation}
\begin{equation}
    \begin{aligned}
        \Delta_U^{[i\sigma^{i+}\gamma_5]}(x,\kv_T) = & \; \frac{1}{M} \epsilon_T^{ij} \kv_{T \; j}H_1^\perp \; , \\
        \Delta_L^{[i\sigma^{i+}\gamma_5]}(x,\kv_T) = & \; \frac{1}{M}S_L \kv_T^i H_{1L}^\perp \; , \\
        \Delta_T^{[i\sigma^{i+}\gamma_5]}(x,\kv_T) = & \; {\bf S}_T^i H_{1T} + \frac{1}{M^2} {\bf S}_T\cdot \kv_T \kv_T^i H_{1T}^\perp \; , \\
        \Delta_{LL}^{[i\sigma^{i+}\gamma_5]}(x,\kv_T) = & \frac{1}{M}S_{LL} \epsilon_T^{ij}\kv_{T \; j} H_{1LL}^\perp \; , \\
        \Delta_{LT}^{[i\sigma^{i+}\gamma_5]}(x,\kv_T) = & \epsilon_T^{ij} {\bf S}_{LT \; j}H_{1LT}^\prime + \frac{1}{M^2} {\bf S}_{LT} \cdot \kv_T \epsilon_T^{ij}\kv_{T \; j} H_{1LT}^\perp \; , \\
        \Delta_{TT}^{[i\sigma^{i+}\gamma_5]}(x,\kv_T) = & \frac{1}{M} \epsilon_T^{ij}S_{TT \; jl} \kv_T^l H_{1TT}^\prime + \frac{1}{M^3} \kv_T \cdot {\bf S}_{TT} \cdot \kv_T \epsilon_T^{ij} \kv_{T \; j} H_{1TT}^\perp \; .
    \end{aligned}
\label{eq: T quark FFs}
\end{equation}

For the gluon FFs, at leading order in $\alpha_s$ there is no $\kv_T$ dependence in the diagrams, and so here we only summarize the FFs which are nonvanishing at this order,
\begin{equation}
    \begin{aligned}
        \Delta_U^{\alpha \beta} = & \; - \frac{1}{2}g_T^{\alpha\beta} D_1^g \; , \\
        \Delta_L^{\alpha \beta} = & \; \frac{i}{2}\epsilon_T^{\alpha \beta} S_L G_1^g \; , \\
        \Delta_T^{\alpha \beta} = & \; 0 \; , \\
        \Delta_{LL}^{\alpha \beta} = & \; - \frac{1}{2}g_T^{\alpha\beta} S_{LL} D_{1LL}^g \; , \\
        \Delta_{LT}^{\alpha \beta} = & \; 0 \; , \\
        \Delta_{TT}^{\alpha \beta} = & \; \frac{1}{2}S_{TT}^{\alpha \beta}H_{1TT}^g \; . \\
    \end{aligned}
\end{equation}

\bibliography{main}

\end{document}